\documentclass[apj,numberedappendix]{emulateapj}
\usepackage{graphics,epsf}
\usepackage{amsmath}                
\usepackage{amsfonts}               
\usepackage{amssymb}                
\usepackage{epsfig}                 
\usepackage{graphicx}
\usepackage{float}
\usepackage{color}
\usepackage[para,online,flushleft]{threeparttable}

\newcommand{\kms}{{~\rm km\; s^{-1}}}

\newcommand{\cm}{{~\rm cm}}
\newcommand{\km}{{~\rm km}}
\newcommand{\s}{{~\rm s}}

\newcommand{\g}{{~\rm g}}

\newcommand{\K}{{~\rm K}}
\newcommand{\erg}{{~\rm erg}}
\newcommand{\yr}{{~\rm yr}}
\newcommand{\Myr}{{~\rm Myr}}
\newcommand{\myr}{{~\rm Myr}}

\newcommand{\kpc}{{~\rm kpc}}

\begin{document}

\title{Uplifted cool gas and heating by mixing in cooling flows}
\author{Shlomi Hillel\altaffilmark{1} \& Noam Soker\altaffilmark{1,2}}

\altaffiltext{1}{Department of Physics, Technion -- Israel Institute of Technology, Haifa 32000, Israel; shlomihi@tx.technion.ac.il; soker@physics.technion.ac.il}
\altaffiltext{2}{Guangdong Technion Israel Institute of Technology, Shantou, Guangdong Province, China}


\begin{abstract}
We analyze our earlier three-dimensional hydrodynamical numerical simulation of jet-inflated bubbles in cooling flow clusters, and find that dense gas that was not heated by the jets' activity and that resides around the hot jet-inflated bubbles can be identified as uplifted gas as observed in some clusters. During the build up of the dense gas around the hot bubble, mixing of hot bubble gas with other regions of the intracluster medium (ICM) heats the ICM.  The vortices that mix the ICM with the hot bubble gas also excite shock waves, sound waves, and turbulence. Sound waves, shocks, turbulence, and uplifted gas, might be easier to detect than the mixing process and hence attract more attention, but we argue that the contributions of these processes to the heating of the ICM do not add up to the level of contribution of the mixing-heating process.
\end{abstract}


\section{INTRODUCTION}
\label{sec:intro}

Jets launched from the active galactic nucleus (AGN) heat the intra-cluster medium (ICM) in cooling flows in clusters, in groups of galaxies, and in galaxies. The heating process of the ICM and the radiative cooling of the ICM operate in negative feedback mechanism (e.g., \citealt{Fabian2012, McNamaraNulsen2012, Farage2012, Gasparietal2013, Pfrommer2013, Baraietal2016}; for a recent review see \citealt{Soker2016}).
Many recent studies support the \emph{cold feedback mechanism} \citep{PizzolatoSoker2005} where cold dense clumps that feed the AGN close the feedback heating mechanism
(e.g., \citealt{Gaspari2015, VoitDonahue2015, Voitetal2015, Lietal2015, Prasadetal2015, SinghSharma2015, Tremblayetal2015, ValentiniBrighenti2015, ChoudhurySharma2016, Hameretal2016, Loubseretal2016, Russelletal2016, McNamaraetal2016, YangReynolds2016b, Baraietal2016, Prasadetal2017, Tremblayetal2016, Donahueetal2017, GaspariSadowski2017, Gasparietal2017, Hoganetal2017, Voitetal2017, Meeceetal2017} to list some papers from the last three years).

While there is a large agreement that cooling and AGN feeding operate via the cold feedback mechanism, there is yet no agreement on the heating process, i.e., the manner by which the jets transfer their energy to the ICM. One can classify the heating processes into two categories. In one type of heating processes the jets and the bubbles they inflate do work on the ICM by exciting shocks (e.g., \citealt{Formanetal2007, Randalletal2015, Guoetal2018}), turbulence (e.g.,    \citealt{DeYoung2010, Gasparietal2014, Zhuravlevaetal2014, Zhuravlevaetal2017}), and sound waves (e.g., \citealt{Fabianetal2006, Fabianetal2017, TangChurazov2018}) , and/or by uplifting gas from inner regions (e.g., \citealt{GendronMarsolaisetal2017}).
In the second type of heating processes energy from the hot jet-inflated bubbles is transfered to the ICM, either as cosmic rays (e.g. \citealt{Fujitaetal2013, FujitaOhira2013}) or by mixing (e.g., \citealt{BruggenKaiser2002,  Bruggenetal2009, GilkisSoker2012,HillelSoker2014, HillelSoker2016, YangReynolds2016b}).
Combinations of two or more processes are also possible, e.g., thermal conduction and cosmic rays (e.g., \citealt{GuoOh2008}), mixing of cosmic rays inside jet-inflated bubbles with the the ICM \citep{Pfrommer2013}, and heating by turbulence and turbulent-mixing (e.g. \citealt{BanerjeeSharma2014}).

Different studies argue that some of the proposed heating processes cannot work efficiently, e.g.,  shocks (e.g., \citealt{Sokeretal2016}) and turbulent heating (e.g., \citealt{Falcetaetal2010, Reynoldsetal2015, Hitomi2016, HillelSoker2017a, Bambicetal2018}). Shocks and turbulence are excited by jet-inflated bubbles, as sound waves are (e.g., \citealt{SternbergSoker2009}), and are indeed observed. Turbulence was detected in cooling flows (e.g., \citealt{Zhuravlevaetal2014, Zhuravlevaetal2015, Arevalo2016, AndersonSunyaev2016, Hofmannetal2016}), most impressive in the the Perseus cluster by the Hitomi telescope \citep{Hitomi2016, Hitomi2017}, but seems to be too weak to account for the required heating of the ICM in cooling flows.

In previous papers we argued that the main heating process by jet-inflated bubbles is mixing of hot bubble gas with the ICM (for a different view see, e.g., \citealt{Weinbergeretal2017}).
The process of bubble inflation by jets forms many vortices on the boundary of the jets and the bubbles with the ICM. These vortices mix the hot bubble gas with the ICM \citep{GilkisSoker2012,HillelSoker2014, HillelSoker2016, HillelSoker2017a, YangReynolds2016b}. The mixing process does not destroy the bubbles and they can buoy out \citep{HillelSoker2014}.
Alongside the heating by mixing, the jet-inflated bubbles excite shocks, sound waves, and turbulence, but these play a smaller role in heating the ICM than mixing does. In the present study we consider recent claims that bubble-uplifted cool gas heat the ICM, and argue that this is another process that comes alongside the inflation of bubbles, but plays a less significant role than heating by mixing.

Uplifting of cool gas by bubbles has been observed and studied for a long time (e.g., \citealt{Russelletal2017, Suetal2017}  for recent papers and references therein). \cite{GendronMarsolaisetal2017} claim that the uplift energy could be an important source of heating in cooling flows. Protrusions that are seen on the boundary of jets and bubbles in the radio image of NGC~1399 (\citealt{Suetal2017}, their figure 4) hint at a complicated interaction process of the jets and bubbles with the ICM, e.g., instabilities. We argue that mixing takes place in the boundary.

We note that there are two basic types of uplifted gas. The first is a hot X-ray emitting gas, but cooler than the surroundings, that forms a bright X-ray rim around the X-ray deficient cavities (bubbles; e.g., \citealt{Russelletal2017, Suetal2017, GendronMarsolaisetal2017}). This is the subject of this paper. The second is cold gas that is observed at much longer wavelength (IR, optical, UV) and it usually trails behind bubbles,
as simulated by, e.g., \cite{Guoetal2018}. As well, \cite{Churazovetal2001} simulated X-ray emitting gas that rails bubbles.

In the present paper we analyze our previous numerical simulations (described in sections \ref{sec:numerics} and \ref{sec:simualtions}) to show that while mixing-heating is taking place on the sides of the jets and bubbles they inflate, a cool gas is uplifted by the bubble (section \ref{sec:results}). In the present study we refer to the uplifting of X-ray emitting gas that surrounds the bubbles.
We discuss and summarize our claims in section \ref{sec:summary}.

\section{NUMERICAL SETUP}
\label{sec:numerics}

We analyze in a new way the same 3D hydrodynamical simulation that we first presented in an earlier paper \citep{HillelSoker2016}. Details of the numerical scheme and of the simulation can be found in that paper. Here we present only the essential properties and parameters of the simulation.
As the simulation contains a large amount of data, we emphasized different aspects in follow-up papers.
We applied our numerical results to the galaxy group NGC~5813 \citep{Sokeretal2016}, to the interpretation of the observations of the Perseus cluster of galaxies by Hitomi \citep{HillelSoker2017a}, and to show that the heating by mixing can account for gentle heating of the ICM \citep{HillelSoker2017b} as argued for by \cite{Hoganetal2017}.

We carried out the 3D hydrodynamical simulation with the {\sc pluto} code \citep{Mignone2007}.
We performed the computation in one octant of space, and took the Cartesian axes in that octant to be $0 \le x \le 50 \kpc$, $0 \le y \le 50 \kpc$ and $0 \le z \le 50 \kpc$. We took $z=0$ to be the symmetry plane, and so the $z$ axis is along the symmetry axis of the jet. In reality the AGN launches two opposite jets, but the octant-grid implies that we study only one jet. At the inner planes of the computational grid, $x = 0$, $y = 0$ and $z = 0$, we enforced reflective boundary conditions. We applied outflow boundary conditions at the outer boundaries of $x = 50 \kpc$, $y = 50 \kpc$ and $z = 50 \kpc$. We did not include viscosity nor heat conduction in the simulation, but we did include radiative cooling.
We used adaptive mesh refinement with a highest resolution of $\approx 0.1 \kpc$.

The properties of the jet are as follows. It has a half-opening angle of $\theta_{\rm j} = 70^\circ$, an initial velocity of $v_{\rm j} = 8200 \kms$, and an initial circular cross section $\sqrt{x^2 + y^2} \leq 3 \kpc$ at the plane $z=0$ where we injected the jet.
We simulate slow and massive jets that are dominated by kinetic energy at injection. We notice that there are numerical simulations that use subsonic  very light jets that are dominated by thermal energy or cosmic rays (e.g., \citealt{GuoMathews2011, Guo2016}).

The jet is intermittent. We injected the jet for a time period of $10 \myr$, starting at time $t=0$, and then shut it off for a quiescence time period of $10 \myr$. The jet-active time intervals are given by
\begin{equation}
20(n-1) \le t_n^{\rm jet} \le 10 (2n-1), \qquad n=1,2,3 \dots .
\label{eq:jet}
\end{equation}

During the active phase the power of the jet and its mass outflow rate are such that for the two jets in reality the power and mass outflow rate are
$\dot E_{2{\rm j}} = 2 \times 10^{45} \erg \s^{-1}$, and
$ \dot{M}_{2{\rm j}} = {2 \dot E_{2{\rm j}}}/{v_{\rm j}^2} = 94 M_{\odot}~\yr^{-1}$, respectively.
Because we injected the jet with a density of about $10^{-26} \g \cm^{-3}$ that is much below the ambient density, and with an initial temperature equal to that of the ambient gas $T_0=3\times 10^7 \K$, the pressure of the jet at injection is much lower than that of the ambient medium. This implies that the initial thermal energy of the jet is much smaller (about 3 per cent) than its initial kinetic energy. Observations (e.g., \citealt{Aravetal2013}) show that such slow, massive, and wide outflows do exist.

At the beginning of the simulation we set the temperature of the ICM to be $T_{\rm ICM} (0) = 3 \times 10^7 \K$ in the entire grid. We included radiative cooling from Table 6 of \cite{SutherlandDopita1993} for a solar metallicity gas.
We took the initial ICM density profile to be (e.g., \citealt{VernaleoReynolds2006})
\begin{equation}
\rho_{\rm ICM}(r) = \frac{\rho_0}{\left[ 1 + \left( r / a \right)
^ 2 \right] ^ {3 / 4}},
\label{eq:den}
\end{equation}
where $\rho_0 = 10^{-25} \g \cm^{-3}$ and $a = 100 \kpc$.
We set the gravitational field to maintain an initial hydrostatic equilibrium, and we kept the gravitational field constant in time.

\section{THE CHARACTARISTICS OF THE SIMULATIONS}
\label{sec:simualtions}

Our simulation that we have presented in four previous papers in the last two years, contains a vast amount of information, and it is impossible to present all aspects in one paper. We here present a new analysis aiming to reveal the formation of the uplifted cool gas, while showing the process of mixing of the hot bubble gas with the ICM. However, the setup of the simulations did not include all the desired features for the present purpose that is motivated by a new observational paper \citep{GendronMarsolaisetal2017}. We therefore first discuss the characteristics and limitations of our simulation.  

In the heavy numerical simulation we have performed we used the simplest initial conditions. One of these simplifications was to take the initial temperature of the ICM to be the same in the entire grid $T_0=3\times 10^7 \K$. Therefore, unlike real clusters of galaxies where the  gas in the inner region is cooler than the ICM gas further out, in our simulation the uplifted gas starts with a temperature that is equal to that of the ICM gas further out. This implies that in our simulation the uplifted gas has about the same temperature as the gas in its surroundings.  

The second feature that does not directly reproduce the cluster NGC~4472 that \cite{GendronMarsolaisetal2017} analyze and that motivated the present study is that we repeat jets activity after a quiescence period of only $10 \Myr$. This has two consequences. The first one is that there is an outward propagating shock ahead of the hot bubble and not far from the bubble. It was formed by the last or next to last jets activity episode. Just behind the shock there is a dense hot gas that is not necessarily present in all clusters. Then there is the shock that is excited by the last jets activity episode, and that eventually breaks out of the bubble. As it breaks out it forms a hot region just above the bubble. Again, this shock does not necessarily exist in all clusters.

\begin{figure*}
\centering
\hskip -0.4 cm
\includegraphics[width=0.50\textwidth]{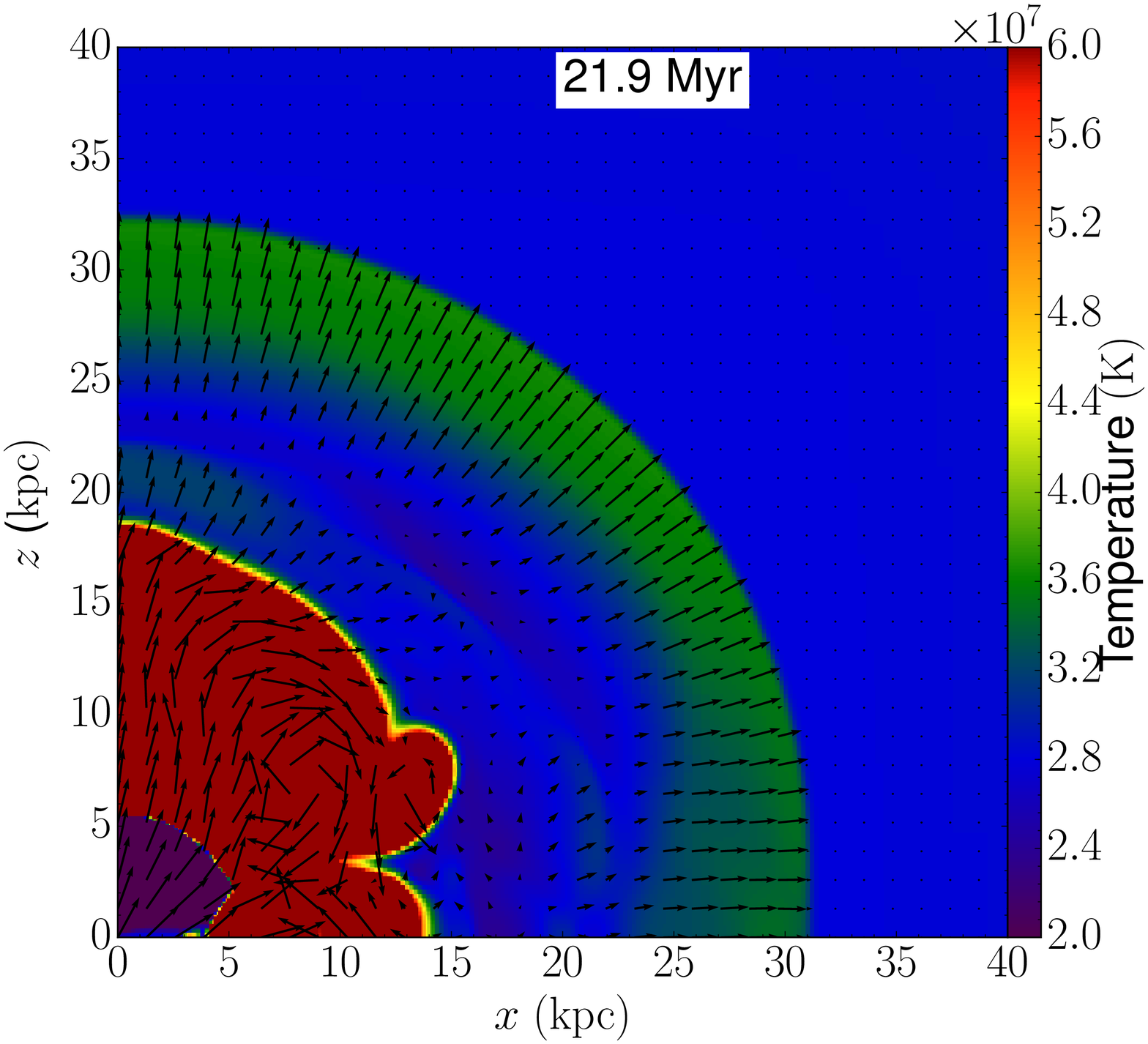}
\hskip -0.2 cm
\includegraphics[width=0.50\textwidth]{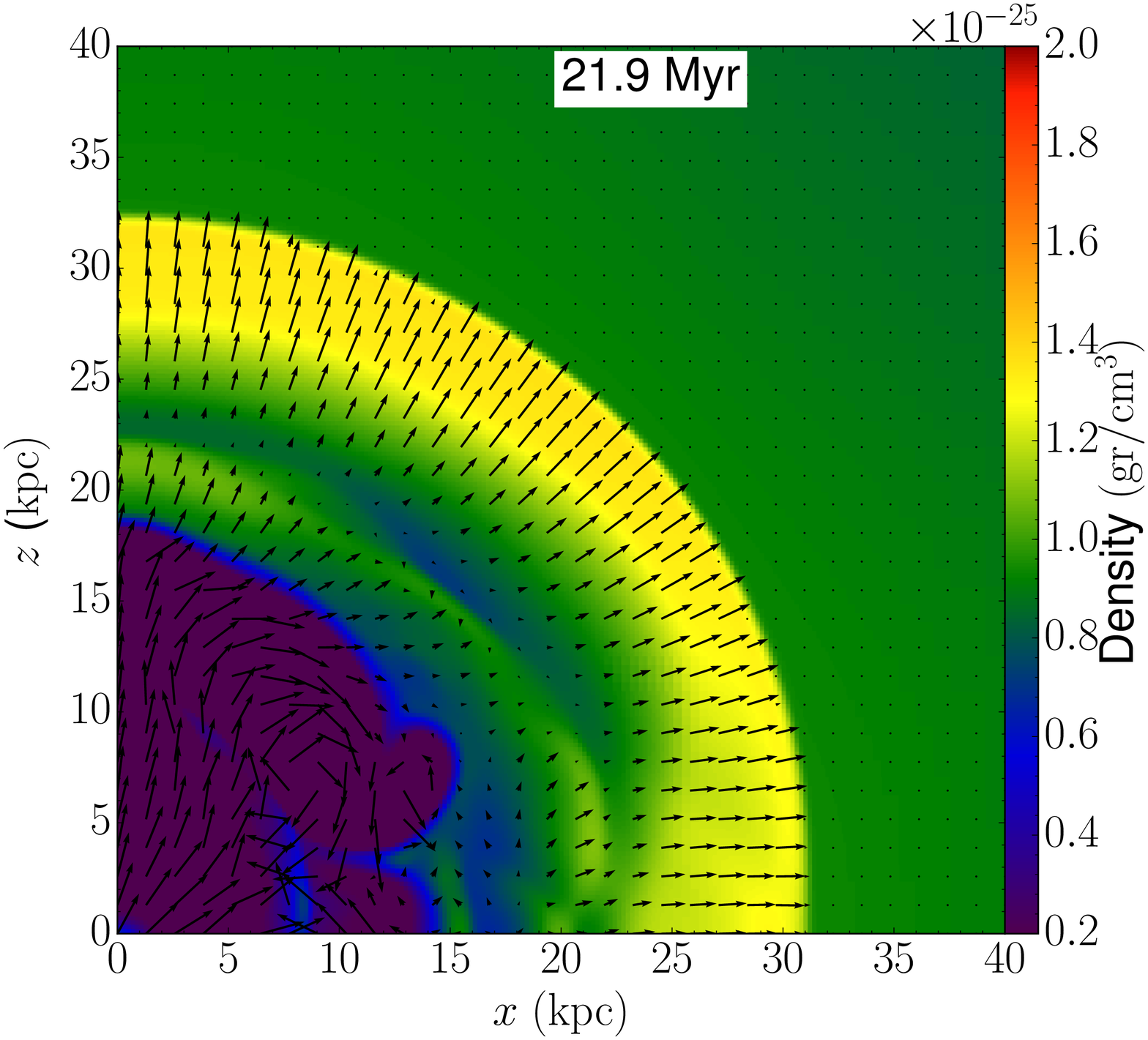} \\
\vskip -0.4 cm
\hskip -0.4 cm
\includegraphics[width=0.50\textwidth]{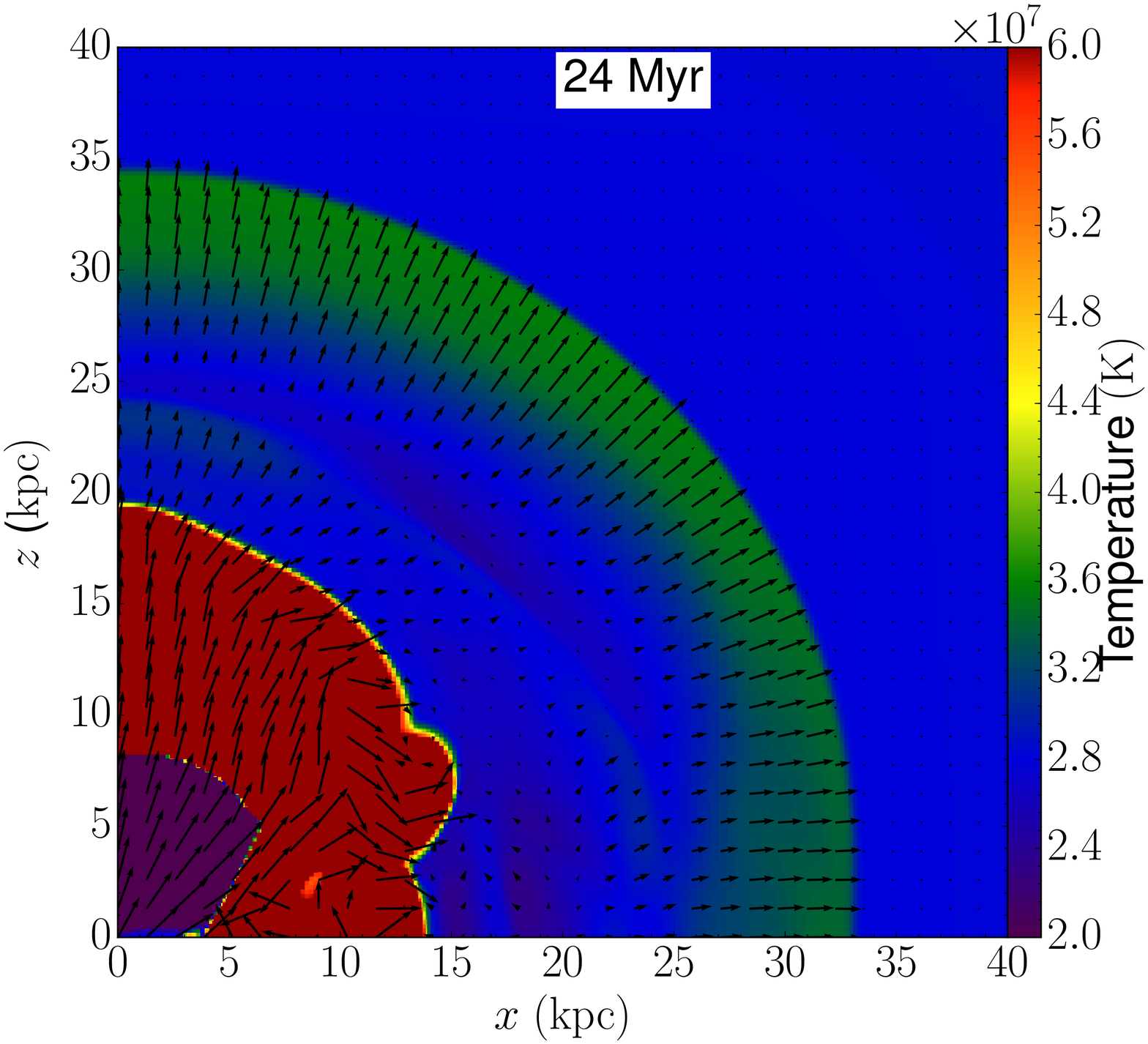}
\hskip -0.2 cm
\includegraphics[width=0.50\textwidth]{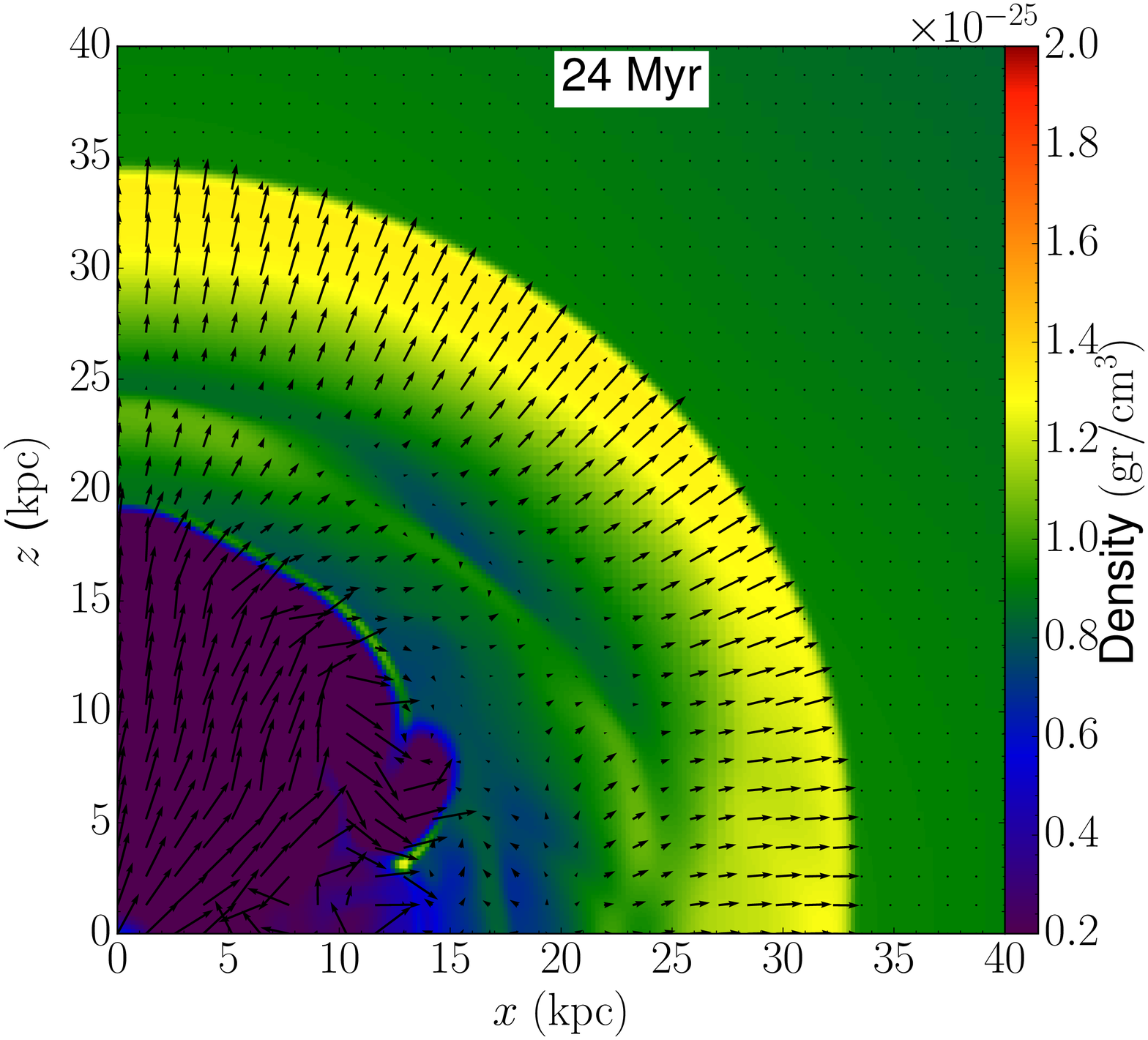} \\
\vskip -0.4 cm
\hskip -0.4 cm
\includegraphics[width=0.50\textwidth]{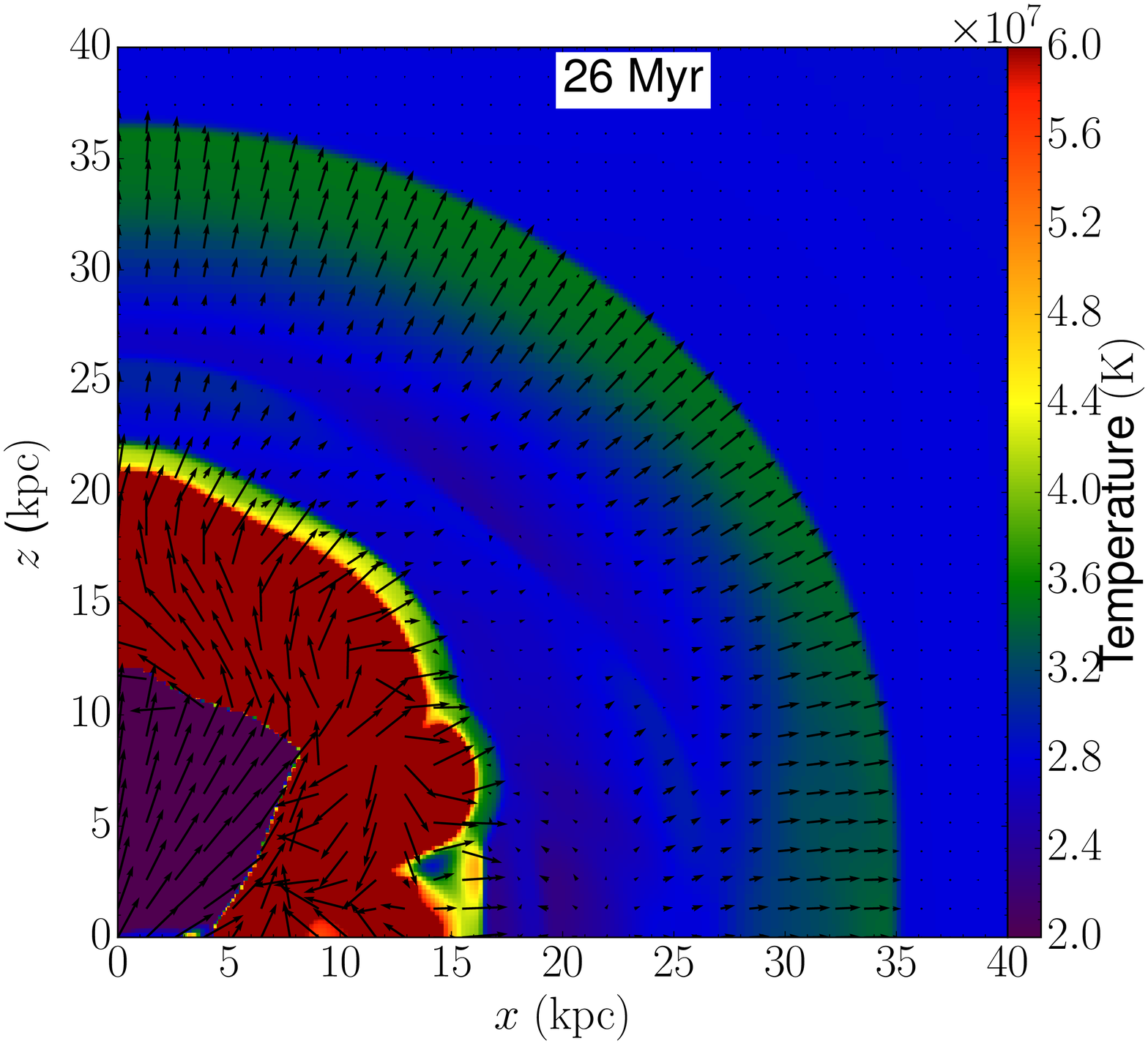}
\hskip -0.2 cm
\includegraphics[width=0.50\textwidth]{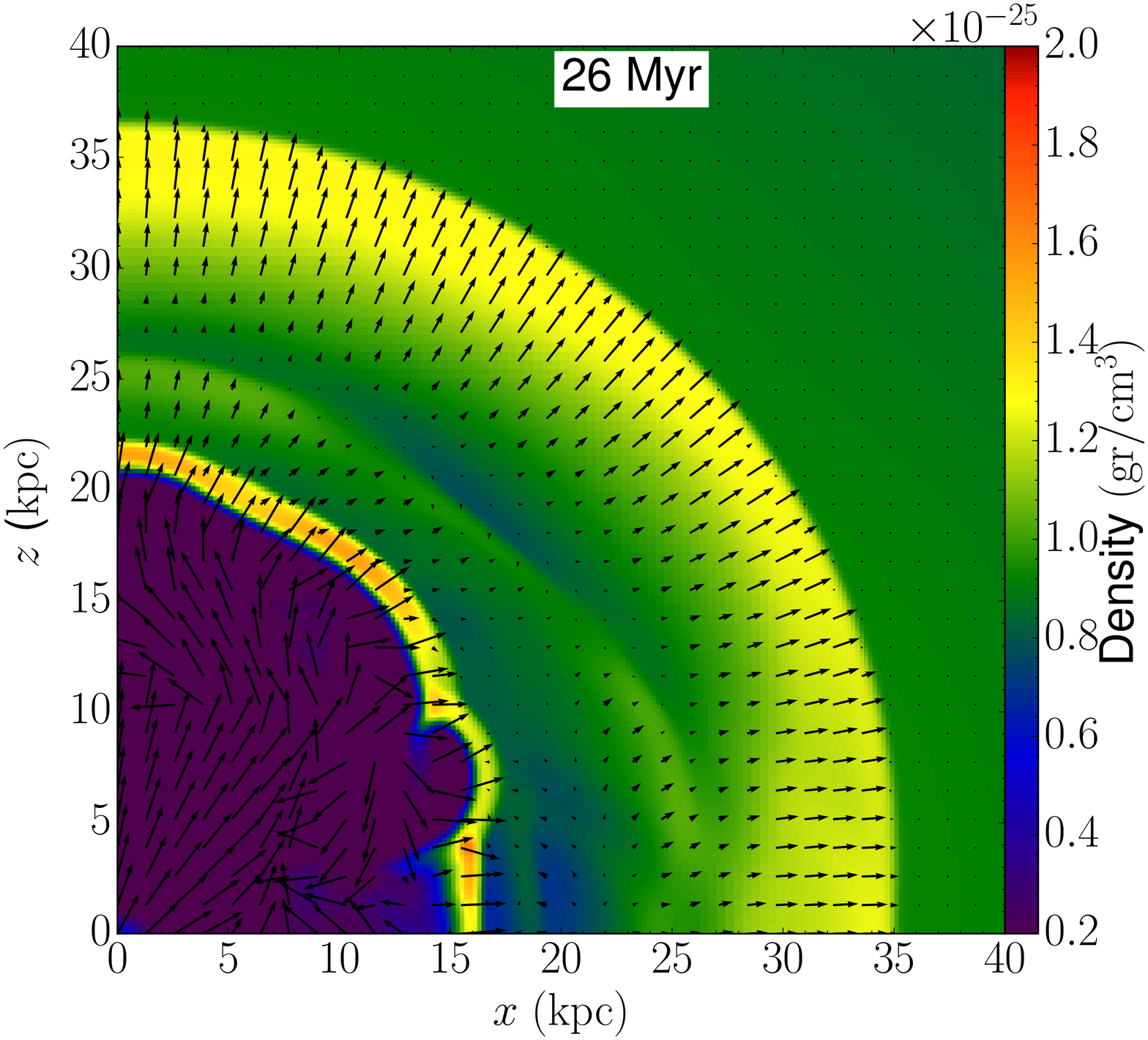}
\vskip -0.4 cm
\caption{Temperature (left) and density (right) maps of the evolution at three times as indicated on the panels after one full jets activity episode and during the second jets activity episode. The temperature and density ranges emphasize the material outside the hot bubble, hence misses very high temperatures and low densities inside the bubble that is seen in red in the temperature maps. The outer shock that was excited by the first episode is propagating ahead of the hot bubble. At $t \simeq 26 \Myr$ the shock that was excited by the second jet-activity episode breaks out from the hot bubble.
The arrows depict velocity with length proportional to the magnitude of the velocity up to $400 \km \s^{-1}$, and higher velocities are marked with arrows with the same length as that of $400 \km \s^{-1}$. Note the vortices close to the equatorial plane. These mix hot bubble gas with the ICM, and lead to mixing-heating.
}
\label{fig:evolution2}
\end{figure*}

We present this evolutionary sequence in figure \ref{fig:evolution2}. Note that the temperature and density colors emphasize the material outside the hot bubbles, hence do not cover the very high temperatures and low densities inside the hot bubble.  For the specific times of the three panels, along the propagation axis (vertical axis) the outer shock propagates from $\simeq 32 \kpc$ to $\simeq 37 \kpc$. The hot post-shock gas is seen in yellow in the temperature maps (left column).
The break out of the next shock is seen in the lower panels as a dense and a hot shell, e.g., as a yellow shell in the temperature map.
If we did not turn on the last jets activity episode, this dense shell would not be there.

\begin{figure*}
\centering
\hskip -0.4 cm
\includegraphics[width=0.50\textwidth]{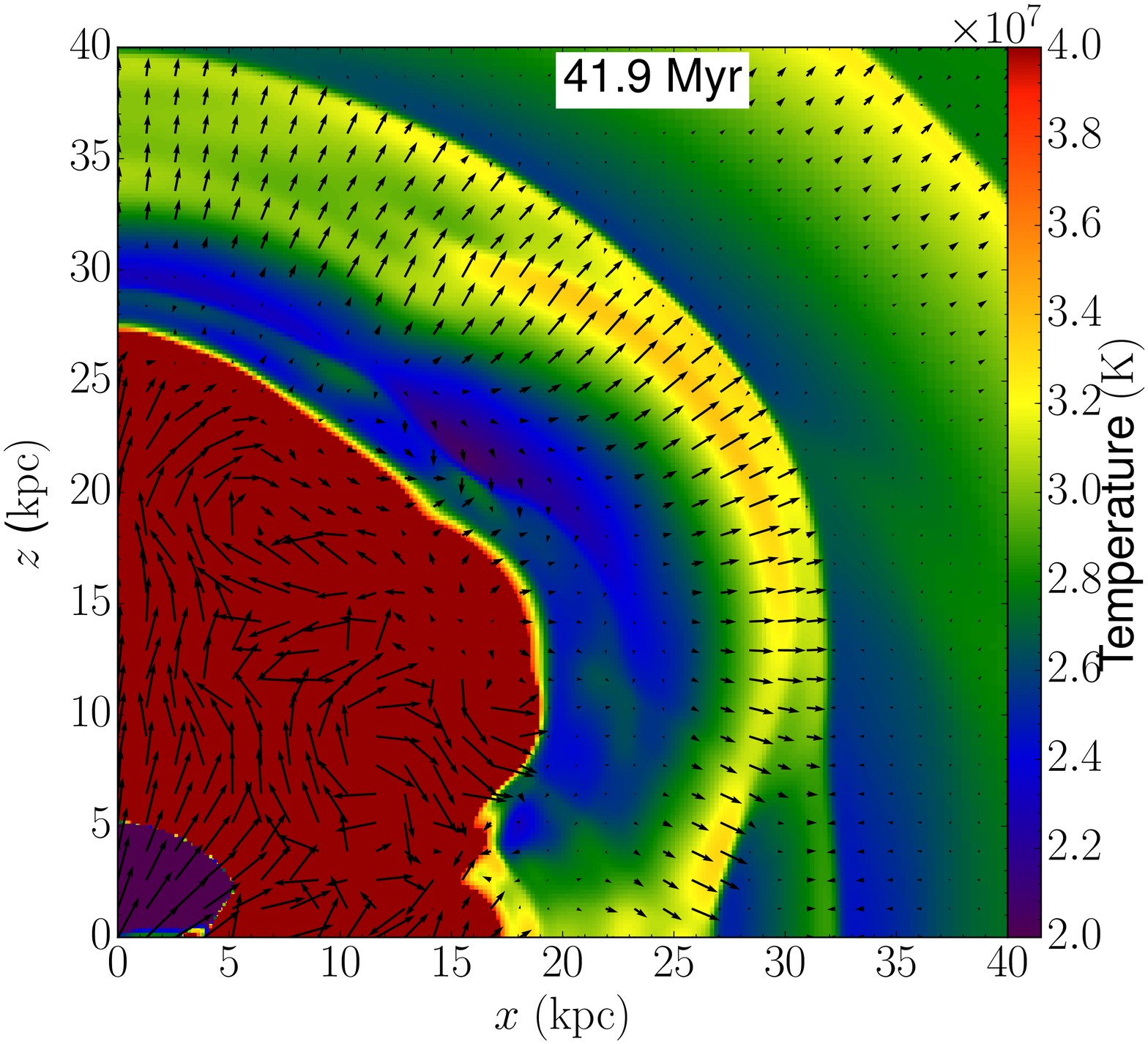}
\hskip -0.2 cm
\includegraphics[width=0.50\textwidth]{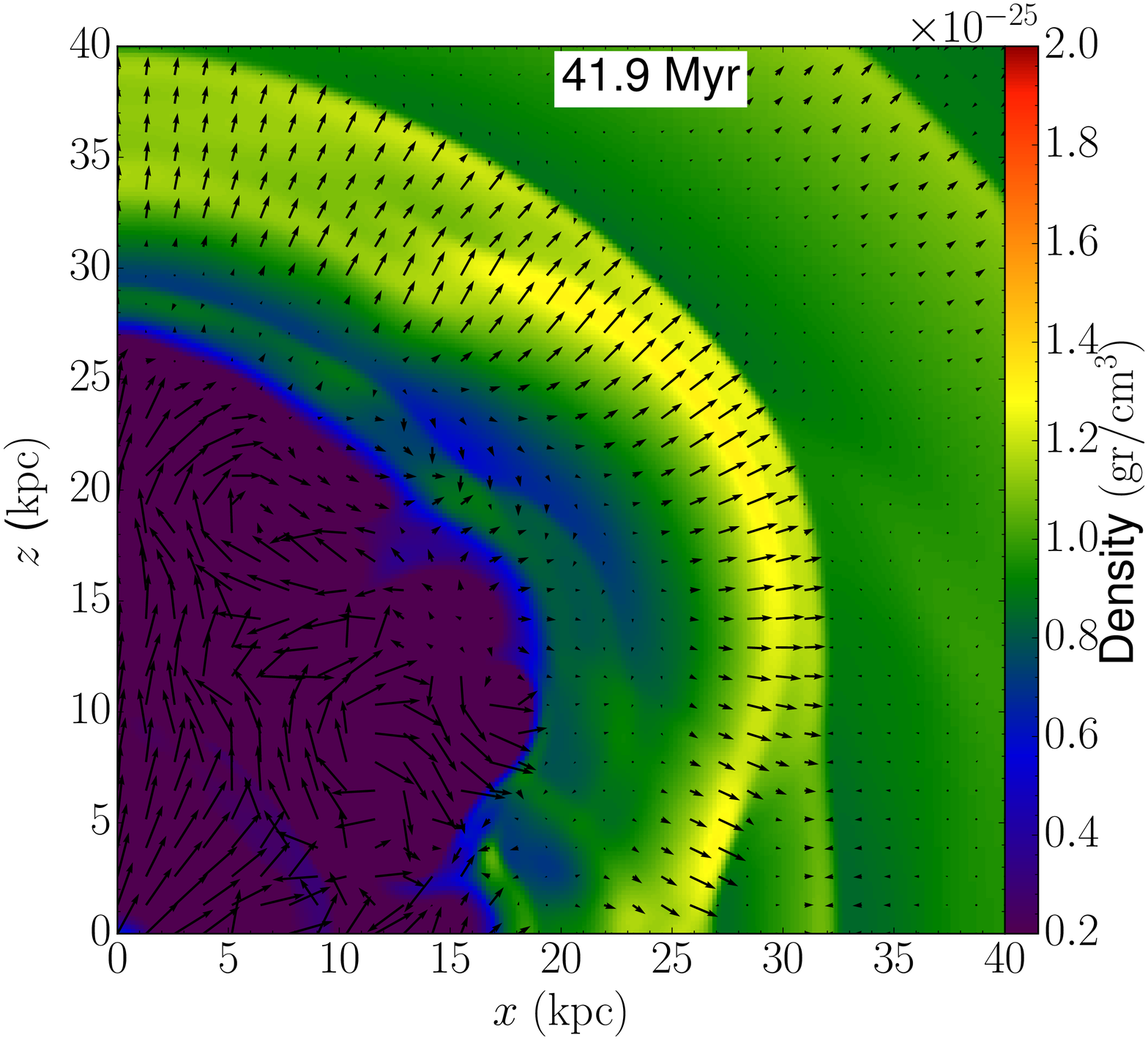} \\
\vskip -0.4 cm
\hskip -0.4 cm
\includegraphics[width=0.50\textwidth]{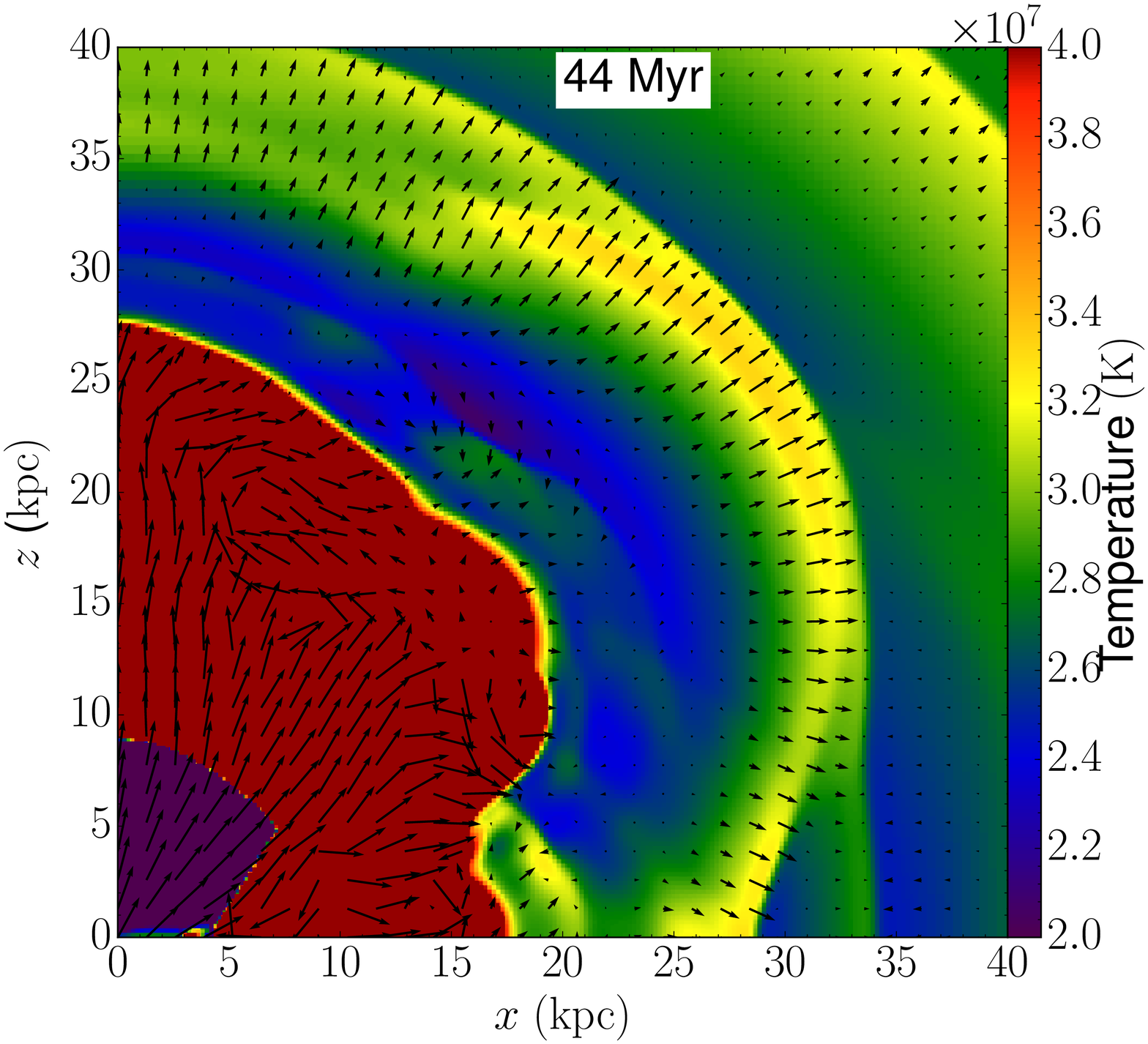}
\hskip -0.2 cm
\includegraphics[width=0.50\textwidth]{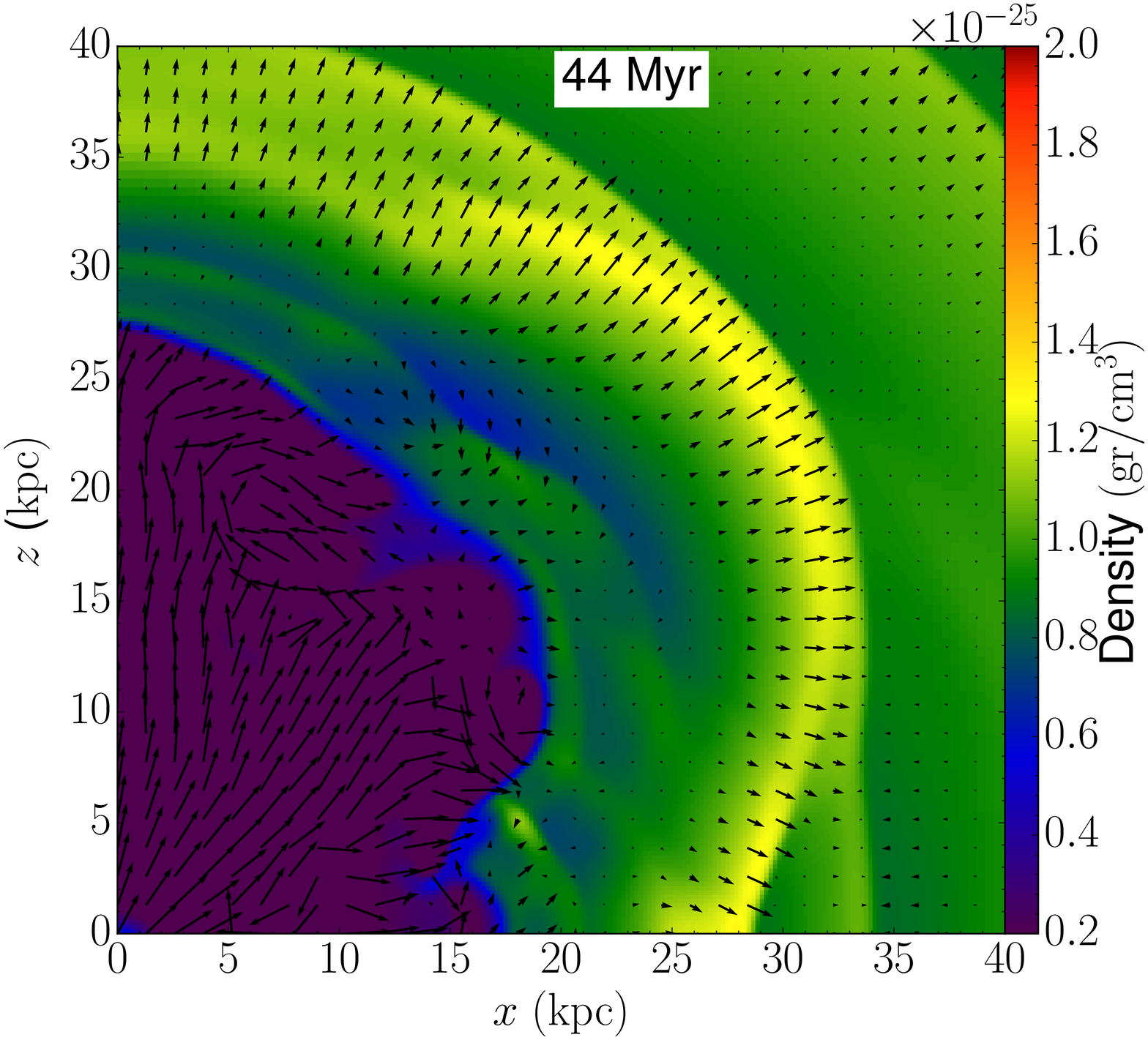} \\
\vskip -0.4 cm
\hskip -0.4 cm
\includegraphics[width=0.50\textwidth]{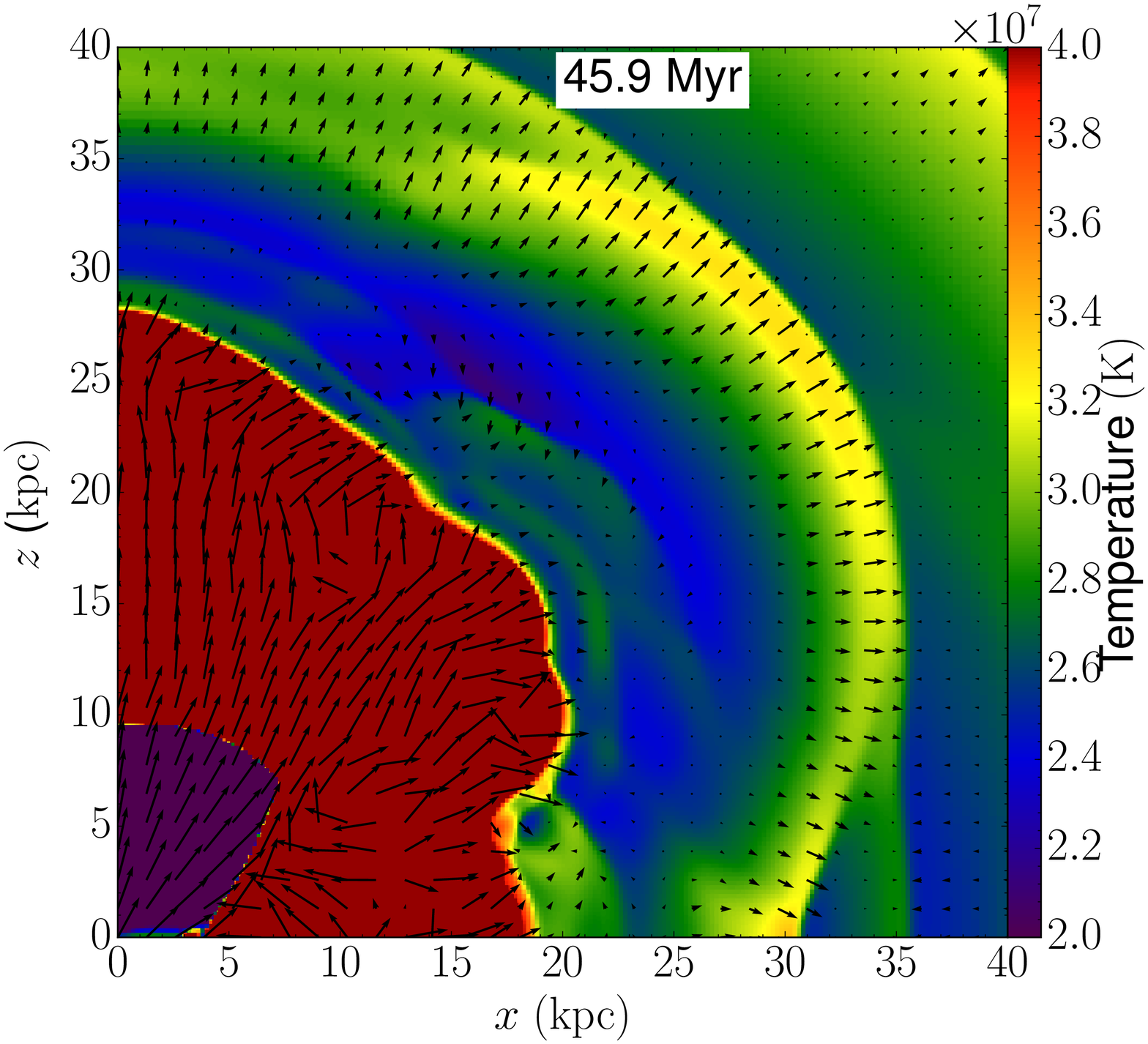}
\hskip -0.2 cm
\includegraphics[width=0.50\textwidth]{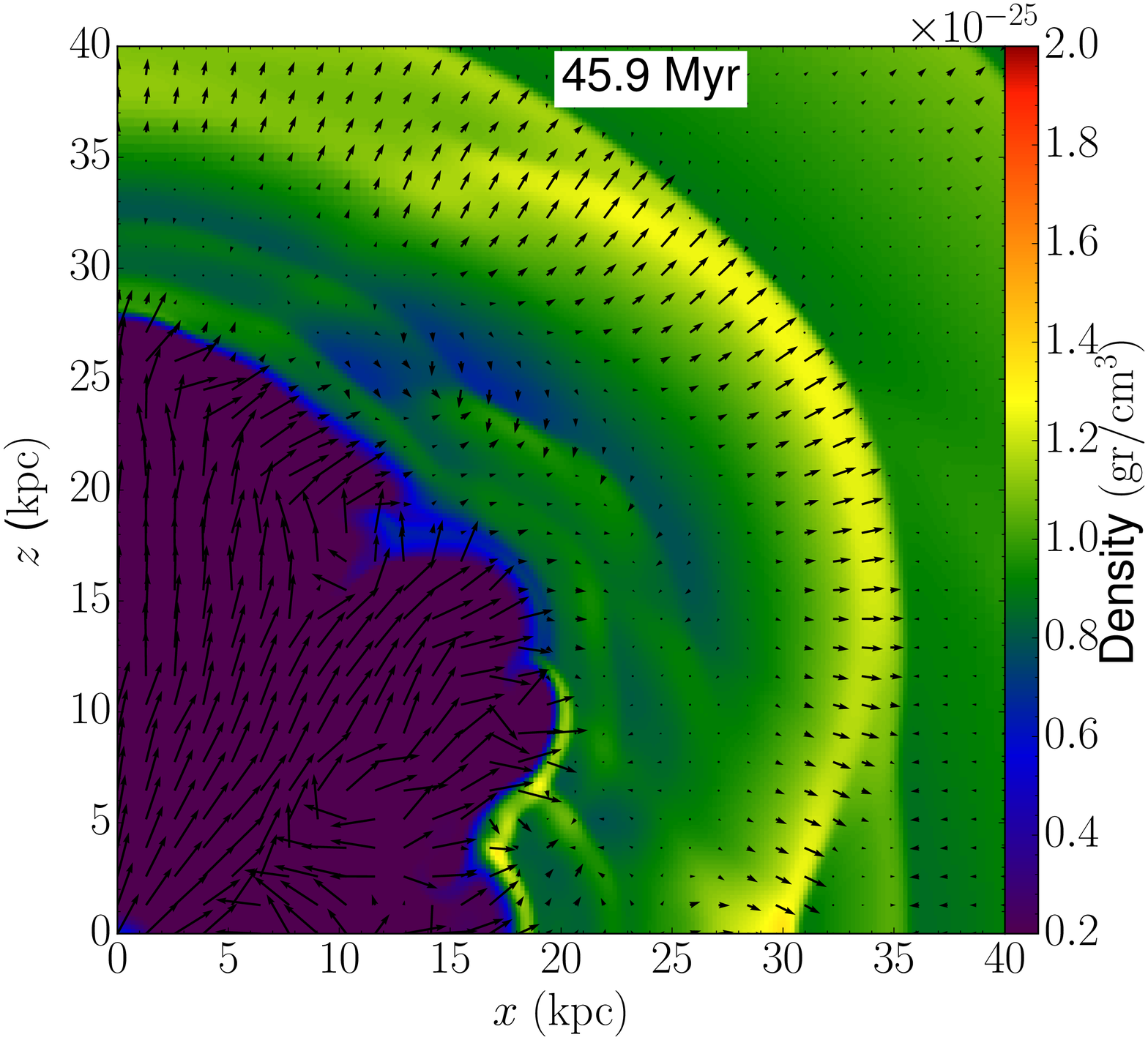}
\vskip -0.4 cm
\caption{Like Fig. \ref{fig:evolution2}, but during the third jets activity episode.
}
\label{fig:evolution3}
\end{figure*}

To further emphasize the characteristics of our simulation, in Fig. \ref{fig:evolution3} we present the evolution while the third jets-activity episode takes place. Now we can see two shocks ahead of the hot bubble, those that were excited by the first two activity episodes.
We can see that after the passage of the second shock a dense layer of gas develops above the hot bubble. This is the uplifted gas that we discuss in the next section.

In what follows, we take it that in the cluster NGC~4472 the last jets-activity episode took place a long time ago, and the shock is far away. We will therefore ignore the appearance of the new shock break-out and the post-shock gas of the shock ahead of the hot bubble, and study only the gas in between the bubble and the outer shock before the next shock break-out.

\section{THE COOL UPLIFTED GAS}
\label{sec:results}

 As seen in Figs. \ref{fig:evolution2} and \ref{fig:evolution3} regions of dense gas develop around the hot bubble after the passage of the last shock. We identify this gas as the uplifted gas that is observed, e.g., in Abell~1795 \citep{Russelletal2017}, in NGC~1399 \citep{Suetal2017}, and in NGC~4472 \citep{GendronMarsolaisetal2017}.
 We mention again that even in the inner regions of our numerical grid the initial temperature of the ICM was $T_0=3 \times 10^7 \K$. What we can see is that the dense gas around the hot bubble was not heated up by the passage of even two shocks. This is along our earlier findings that the main heating process of the ICM is mixing of hot bubble gas with the ICM \citep{HillelSoker2016}. The density of what we identify as the uplifted gas is about $5 \%$ higher than the ambient density (density ratio of $\approx 1.05$). Had we started with a realistic cooler core, the density ratio would be higher than $1.05$.

\begin{figure}
\centering
\includegraphics[width=0.55\textwidth]{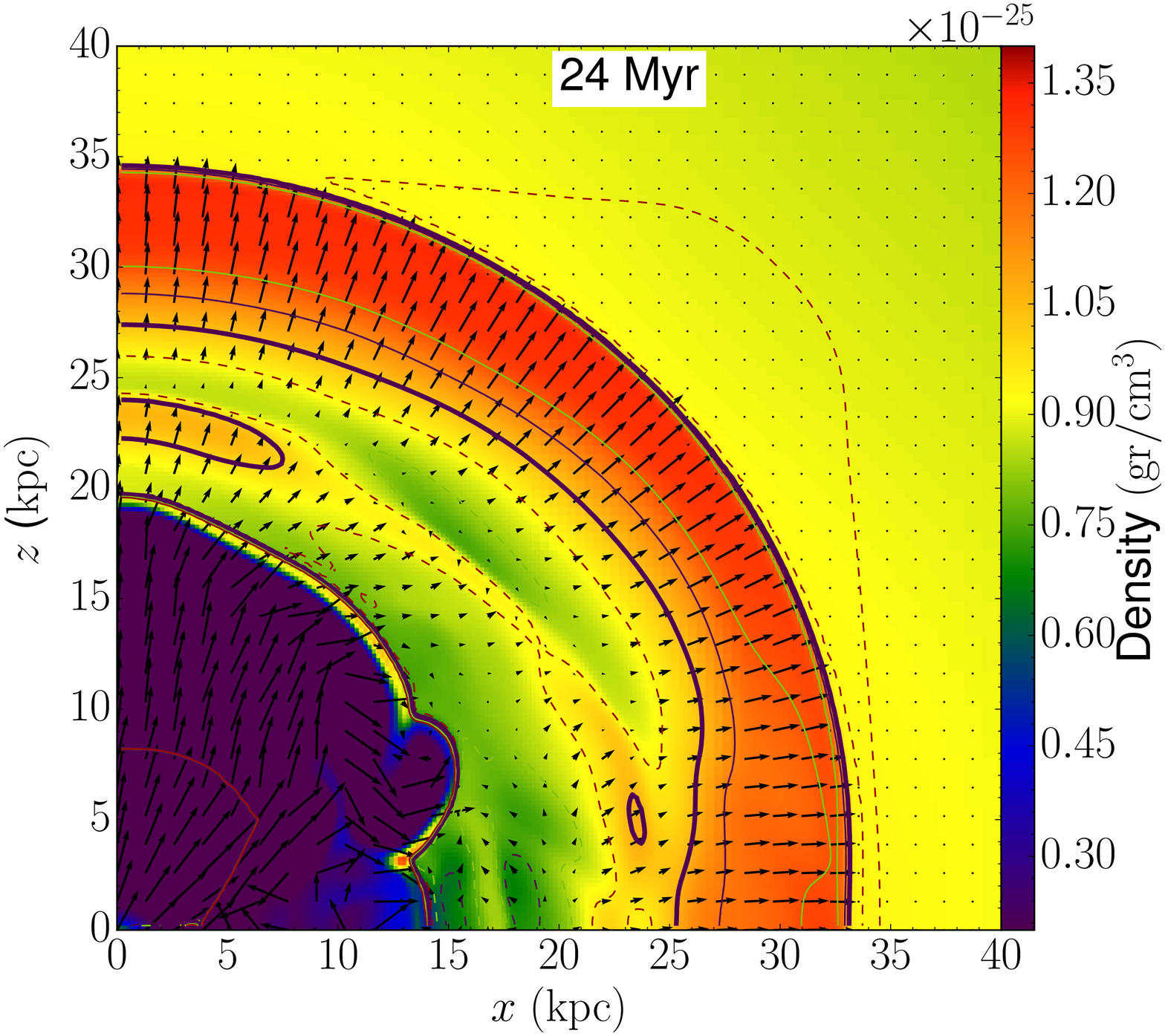}
\includegraphics[width=0.55\textwidth]{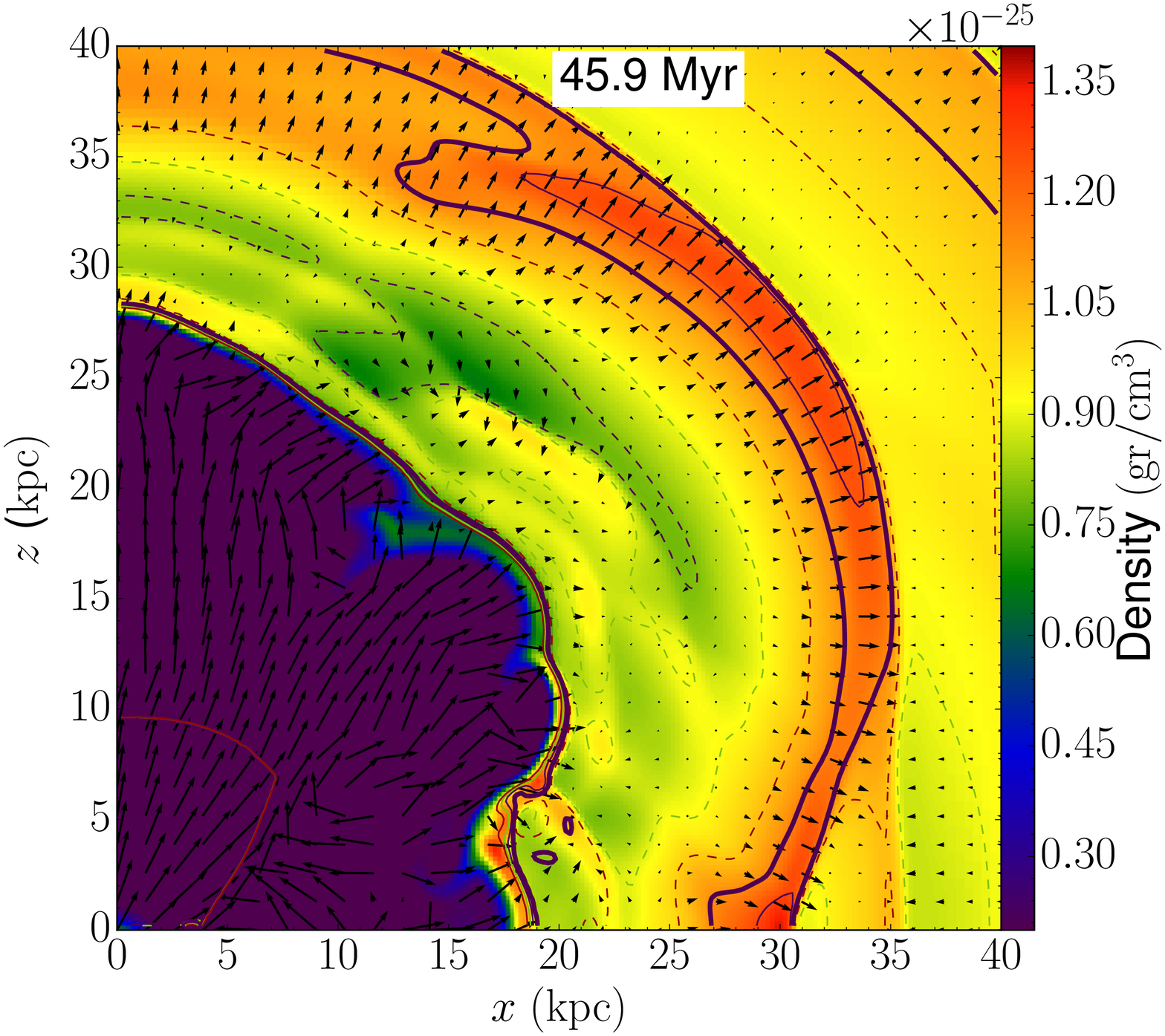}
\caption{ Density maps at two times in the simulation, with superimposed constant-temperature contour lines. The thick-solid line corresponds to $T_0 = 3 \times 10^7 \K$ which is the initial temperature of the ICM. Thin-dashed lines correspond to lower temperature of $T = \{2.4, 2.6, 2.8\} \times 10^7 \K$, and thin solid lines correspond to higher temperatures of $T = \{3.2, 3.4, 3.6\} \times 10^7 \K$. }
\label{fig:contours}
\end{figure}

To further demonstrate that the dense regions around the bubble are not hotter than their initial temperature, in Fig. \ref{fig:contours} we present temperature contours superimposed on the density color maps at two times.

  Let us examine the region around the coordinates $(x,z)=(8 \kpc, 20 \kpc)$ in the upper panel of Fig. \ref{fig:contours}. Closer to the vertical axis there is a dense region with a temperature just slightly above the initial temperature (closed with the thick line). Then there is a region between the thick temperature contour line and the contour line of $T=2.8\times 10^7 \K$. This dense gas is cooler than its initial temperature.
  Had we started with a realistic cooling flow temperature profile, where the gas in the center is cooler, the region we discuss would have been cooler and, because of pressure equilibrium, denser than the present gas there. The same holds for some other regions around the hot bubble.

We can identify similar regions around the hot bubble at later times in the lower panel of Fig. \ref{fig:contours}. For example, there is an elongated dense and low temperature gas along $x=22 \kpc$ in the range $11 \kpc \la z \la 21 \kpc$.

  We can see dense gas that is hotter than the initial temperature at two types of regions (red in the two panels of Figs. \ref{fig:contours}). The first type of region is the gas immediately behind the outer shock(s). But this gas then expands and cools adiabatically and its entropy does not increase much \citep{GilkisSoker2012, HillelSoker2014}.
  The second type of heated gas can be seen on the outskirts of he bubble near the equatorial plane. This is ICM gas that is being heated by its mixing with the hot bubble gas. This, we argue, is the main heating process of the ICM.

\section{DISCUSSION AND SUMMARY}
\label{sec:summary}

Uplifted (dragged) dense gas is clearly seen in numerical simulations of jet-inflated bubbles, e.g., \cite{Sternbergetal2007}, figure 1 in \cite{SternbergSoker2008} and figure 5 of \cite{GilkisSoker2012}. We here described dense gas that was not heated and that resides around the hot bubble, as appears in our earlier 3D hydrodynamical simulation of jet-inflated bubbles (section \ref{sec:results}). We identify this dense gas as the observed uplifted gas.

From their observations of the cluster NGC~4472 \cite{GendronMarsolaisetal2017} argue that when the uplifted gas falls back toward the center it can heat the ICM.
We note the following. When the dense gas falls back it moves subsonically as its density is not much larger than that of the ambient gas. This relative motion pushes gas and excites sound waves. The sound waves are expected to carry energy out of the cooling flow region.
The falling dense gas can excite turbulence, but as Hitomi showed for the Perseus cluster \citep{Hitomi2016, Hitomi2017}, the turbulence is too weak to account for heating.

If the bubbles lift gas that later falls back, then the same amount of energy, or even more, is expected to dissipate as the dense gas is dragged out, because the velocity out is not much different than the fall-back velocity toward the center.

In the observations of NGC~4472 \citep{GendronMarsolaisetal2017} the dense gas was lifted a distance of about $4 \kpc$ from the center in about $18 \Myr$. The average outward velocity is about $200 \km \s^{-1}$.
The fall-back velocity is a fraction of the sound speed, and is not expected to be much larger, or even not larger at all, than this value \citep{PizzolatoSoker2010}. Overall, the fall back can add to the turbulence of the ICM, which at least in the case of the Perseus cluster that was observed with Hitomi \citep{Hitomi2016} is not sufficiently strong to heat the ICM \citep{Fabianetal2017}.

Over all, the motion of the dense clumps and their distortion seems to perturb the ambient medium and excite weak sound waves. This can be seen by the temperature fluctuations around the clumps in a simulation that we presented in an earlier paper (figure 10 in \citealt{HillelSoker2016}). We argue that when dense clumps fall-in, with velocity much lower than the sound speed, sound waves carry most of the gravitational energy that is released by the clumps. The sound waves dissipate in the large volume of the cluster and do not heat much the inner zones where the gas cools the most.

The main result of the present paper is that we show that dense gas, that is identified as uplifted gas, is formed  around hot bubbles as they are inflated and rises outward, while mixing heats the ICM.
Although we attribute most of the heating to mixing, uplifted gas does play a crucial role in the feedback cycle. \cite{PizzolatoSoker2005} argue that
jets and bubbles form nonlinear perturbations that later form to cold gas condensation that feed the AGN within the frame of the cold feedback mechanism. \cite{Hoganetal2017} further suggest that uplifted gas is likely to be the seed of these nonlinear perturbations. The uplifted gas that is later accreted by the AGN can cause the directions of later jets to substantially change, hence leading to jittering jets in cooling flows
\citep{Soker2018}. Such jittering jets make the heating process of the ICM by mixing more uniform.

On a broader scope, we argue that the main heating process of the ICM is mixing of hot bubble gas with the ICM. This mixing is caused by many vortices that are formed during the inflation process of the bubbles. The vortices also excite sound waves, shocks, and turbulence. The inflation of the bubbles and their outward motion further excites sound waves, turbulence, and now we have shown that dense gas forms around the bubbles. These effects, of sound waves, shocks, turbulence, and uplifted gas, might be easier to detect than the mixing process and hence attract more attention.
But all these processes that accompany the inflation of bubbles do not contribute to heating as much as the mixing-heating contributes \citep{HillelSoker2016, HillelSoker2017a}.

We thank an  anonymous referee for helpful comments.
This research was supported by the the Prof. A. Pazy Research Foundation, by the Israel Science Foundation, and by the E. and J. Bishop Research Fund at the Technion.

\label{lastpage}
\end{document}